\documentclass[9pt]{article}
\usepackage{amsfonts}
\usepackage{pstricks}
\usepackage{pst-node}
\usepackage{epsfig}
\usepackage{epsfig}
\usepackage{graphicx}
\usepackage[font={footnotesize,it}]{caption}
\usepackage[latin5]{inputenc}
\usepackage[T1]{fontenc}
\usepackage{lipsum}
\usepackage{amsmath}

\begin{document}
                \def\ba{\begin{eqnarray}}
                \def\ea{\end{eqnarray}}
                \def\w{\wedge}
                \def\d{\mbox{d}}
                \def\D{\mbox{D}}

\begin{titlepage}
\title{Scattering of massive neutrino test fields from a gravitational pulse}
\author{Tekin Dereli${}^{1,2}$\footnote{tekin.dereli@ozyegin.edu.tr}, Yorgo Senikoglu${}^{3}$\footnote{ysenikoglu@gsu.edu.tr}}
\date{%
    ${}^{1}$ \small Professor of Physics, Faculty of Aviation and Aeronautical Sciences, \"Ozye\~gin University, 34794 
    \c{C}ekmek\"{o}y, \.{I}stanbul, Turkey\\%
    ${}^{2}$ \small Department of Physics, Ko\c{c} University, 34450 Sar{\i}yer, \.{I}stanbul, Turkey\\
    ${}^{3}$ \small D\'{e}partement de Math\'{e}matiques, Universit\'{e} Galatasaray, 34349 Be\c{s}ikta\c{s}, \.{I}stanbul, Turkey\\
    \today
}

\maketitle



\begin{abstract}
\noindent Linearized Einstein-Weyl equations are solved precisely in the context of sandwich gravitational waves. The neutrino's energy-momentum depends on the geometry and composition of the gravitational pulse when it is scattered. Since the background remains unchanged at the test field level, the neutrino's energy density will exhibit fluctuations between positive and negative extremes when traversing the sandwich wave. These variations could provide insights into the behavior of models concerning neutrino oscillations.
\end{abstract}

\vskip 1cm


\end{titlepage}

\newpage

\section{Introduction}
\noindent The conjecture of gravitational waves is one of the most fascinating realization of the general theory of relativity. For decades, however, many physicists doubted whether their disclosure would ever be possible with a direct experimental confirmation \cite{rosen},\cite{einstein-rosen}. Despite these doubts, theoretical research into the properties of gravitational waves thrived \cite{brinkmann},\cite{peres}. Among the exact solutions to Einstein's field equations, gravitational plane waves proved particularly enlightening, exposing notable effects such as focusing, nonlinear interaction and formation of curvature singularities \cite{ehlers-kundt} - \cite{griffiths0}. 

\noindent With the direct detections done by the Advanced LIGO and Advanced VIRGO observations, the subject is not only theoretical but has become a pedestal for observational astrophysics \cite{abbott et al}. Unlike electromagnetic radiation, which interacts linearly with matter, gravitational wave phenomena are nonlinear such as the mergers of compact objects. This matter makes it exceptionally fascinating to wonder how fields respond whey they traverse this gravitational wave background \cite{gibbons}. Solar and electromagnetic fields in such backgrounds have been extensively analyzed, while neutrino fields, especially endowed with mass, remain less explored. Neutrinos are specially appealing to study in this context, because their weak interactions make them exemplary cosmic emissaries. The discovery of neutrino oscillations has established that they possess a small nonzero mass. Earlier works on massless neutrinos in plane wave backgrounds \cite{brill-wheeler} - \cite{griffiths1} and their collisions in these geometries have been addressed thoroughly \cite{griffiths2} - \cite{dereli-tucker}, but the introduction of mass may change the dynamics. Later developments also included the propagation of Dirac fields through plane gravitational waves \cite{bini-ferrari2}, where it was shown that the scattering alters the wave function without affecting the spin state. More recent analyses, considered Dirac particles in pure gravitational sandwich waves \cite{collas}, demonstrating that such backgrounds can modify the initial spin polarization of the particles.  These findings are consistent with earlier investigations \cite{halilsoy},\cite{albadawi-halilsoy} into the global form of sandwich waves in Einstein and Einstein-Maxwell theories, which established that test scalar particles may extract energy when traversing such spacetimes. It is important to stress that in all these cases, the scalar and Dirac fields were treated as test fields: the gravitational background itself was fixed, and only the response of the matter fields to the curved geometry was examined.

\noindent A gravitational sandwich wave is a localized burst of curvature with a short lived, extreme astrophysical phenomena such as black hole or neutron star mergers. One may expect neutrinos to undergo changes in their polarization when entering such regions. In test field treatments, we have shown in a previous paper \cite{dereli-senikoglu} that significant variations in energy distributions across sandwich waves may be observed raising the question whether such effects could enhance, modify or cause neutrino oscillations between different flavors. In the present work, we turn to the case of massive neutrino fields propagating through sandwich plane waves. Our analysis will focus on massive test neutrinos in three types of sandwich wave backgrounds: (i) purely gravitational, (ii) purely electromagnetic, and (iii) mixed gravitational-electromagnetic configurations. By extending the discussion beyond the massless case, we aim to clarify how the small but finite mass of neutrinos reshapes their propagation properties in curved plane wave geometries.

\noindent The structure of the paper is organized as follows: Section 2 provides a general overview of sandwich gravitational waves in Einstein-Maxwell theory. In Section 3, we solve the massive neutrino field equations in the context of the sandwich gravitational wave background. Section 4 presents detailed calculations of energy variations encountered when crossing the sandwich gravitational waves. Finally, Section 5 offers the conclusions of our study.

\bigskip

\section{Gravitational Sandwich Waves}
The Brinkmann\cite{brinkmann} representation of the gravitational or electromagnetic pp-wave metric is given by:
\begin{equation}\label{Brinkmann}
g=2dUdV + 2 H(U,X,Y)dU^2 + dX^2 + dY^2.
\end{equation}

\noindent A natural generalization of shock or step waves, known as sandwich waves, can be characterized using Heaviside step functions, which remain nonzero within a finite interval $0\leq U \leq U_0$. In this context, both the region preceding the wavefront $U=0$ and the area following it at $U=U_0$ are described by the flat Minkowski metric. Within the framework of Einstein-Maxwell theory, a general form of the sandwich plane wave metric is discussed in \cite{halilsoy}. By imposing suitable constraints, one can construct purely gravitational waves, purely electromagnetic waves, or mixed configurations that remain valid within the finite-duration curvature region. Notably, the gravitational component $\psi_4$ and the electromagnetic component $\phi_{22}$ of a linearly polarized plane sandwich wave are independent of the transverse coordinates $X$ and $Y$. The metric function in this case becomes:
\begin{equation}
  H(U,X,Y)=\frac{1}{2}[(\Theta(U)-\Theta(U-U_0)][a^2(X^2+Y^2)-b^2(X^2-Y^2)],
\end{equation}
where the specific constants $a$ and $b$ are representative of the electromagnetic and the gravitational framework, respectively.

It is often more convenient to use Rosen's metric formulation \cite{rosen} to explicitly illustrate the transverse structure of these spacetimes through the following coordinate transformation:
\ba
U=u, \quad U_0=u_0, \nonumber \\ 
V=v + \frac{1}{2}(x^2FF_u+y^2GG_u),\nonumber \\ 
X=xF, \quad Y=yG
\ea
so that the metric becomes
\begin{equation}\label{Rosen}
  g=2dudv + F(u)^2 dx^2 + G(u)^2 dy^2.
\end{equation}
The functions $F(u)$ and $G(u)$ are governed by the following system of differential equations:
\ba
F_{uu}+A^2(\theta(u)-\theta(u-u_0))F=0, \nonumber \\
G_{uu}+B^2(\theta(u)-\theta(u-u_0))G=0.
\ea
The general solution is expressed as:
\ba \label{eq9}
F(u)=cos[A(u\theta(u)-(u-u_0)\theta(u-u_0))]-Asin(Au_0)(u-u_0)\theta(u-u_0), \nonumber \\
G(u)=cos[B(u\theta(u)-(u-u_0)\theta(u-u_0))]-Bsin(Bu_0)(u-u_0)\theta(u-u_0).
\ea
We set $A^2=(a^2-b^2)$ and $B^2=(a^2+b^2)$.
In this paper, we analyze the gravitational wave background in three distinct cases: first, a purely gravitational wave; second, a purely electromagnetic wave; and third, a combination of gravitational and electromagnetic sandwich waves. In each scenario, the relevant physical quantities can be derived from the general solution by taking the appropriate limiting cases.
The overall framework for studying this gravitational sandwich wave geometry is depicted in Figure 1.

\begin{figure}[htb!]
  \centering
  \includegraphics[width=1.00 \textwidth]{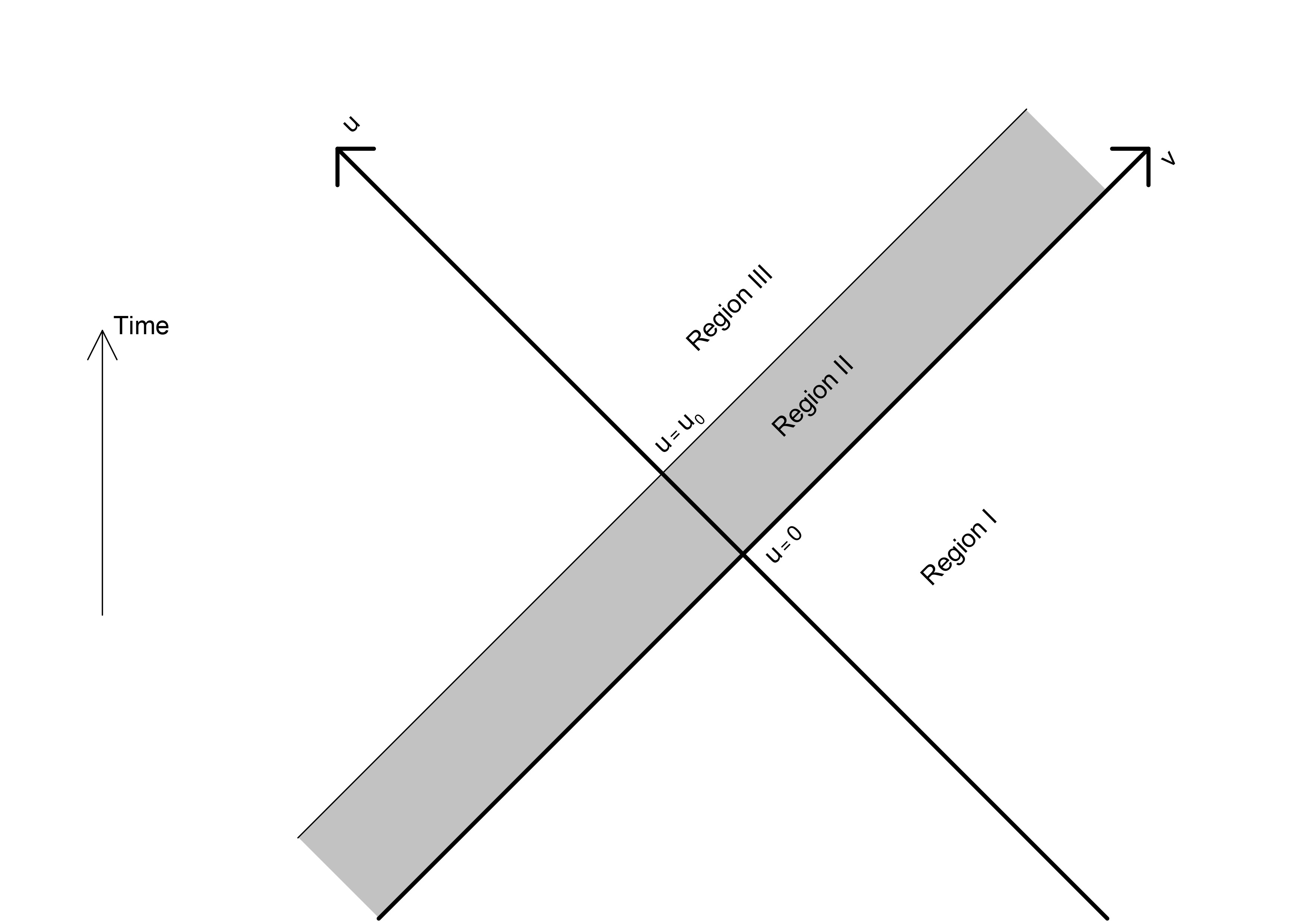}
  \caption{The general structure of the gravitational sandwich wave geometry consists of three regions. Region I represents the flat Minkowski spacetime ahead of the sandwich wave. Region II is the curved domain where the sandwich wave is present. Region III corresponds to the flat Minkowski region trailing the wave.}
\end{figure}

\noindent The three distinct regions-two flat regions on either side and a curved region within the finite duration plane fronted wave-must be smoothly joined using appropriate junction conditions. Previous studies \cite{halilsoy} have demonstrated that the O'Brien and Synge junction conditions are satisfied across the boundaries at $u=0$ and $u=u_0$ for the global sandwich wave solution in Einstein-Maxwell theory.
\bigskip

\section{Massive Neutrino Equation in a Sandwich Wave}
Before making any assumptions, a clarification must be made on how to assign a mass to a spinor field. One possible way to introduce a mass term is to consider the following coupled system of equations satisfied by 2-component spinors $\phi$ and $\chi$:
\ba
\sigma^{a} \nabla_{X_{a}} \phi &=& m_D \dot{\chi}, \nonumber \\ 
\bar{\sigma}^{a} \nabla_{X_{a}} \dot{\chi} &=& m_D \phi.
\ea
Here, $\sigma^a :\{I ,\sigma^1, \sigma^2, \sigma^3 \}$ represent the Pauli matrices, while $\nabla_{X_{a}}$ denotes the covariant derivatives with respect to an orthonormal
co-frame $\{e^a\}$. Consequently the covariant exterior derivative operator is given by $\nabla = e^a \nabla_{X_{a}}$. 
The latter equations are nothing but another way of writing out the Dirac equation
\ba
\gamma^a \nabla_{X_{a}} \Psi = m_D \Psi ,
\ea
where we have 
\ba
\Psi = \left ( \begin{array}{c} \phi \\ \dot{\chi} \end{array}  \right ), \quad 
\gamma^a=
\left( {\begin{array}{cc}
0 & \bar{\sigma}^a \\
\sigma^a & 0\\
\end{array} } \right),
\ea
and $m_D$ is the Dirac mass.

\smallskip
\noindent
Nevertheless, there is another way to introduce a mass term, let us consider
\ba
\sigma^{a} \nabla_{X_{a}} \Phi =m \dot{\Phi}.
\ea

\noindent
This implies and is implied by the conjugate equation
\ba
\bar{\sigma}^{a} \nabla_{X_{a}} \dot{\Phi} =m \Phi.
\ea
We iterate either one of these equations and get the Klein-Gordon equation:
\ba
\bar{\sigma}^{a} \nabla_{X_{a}} (\sigma^{b} \nabla_{X_{b}} \Phi) &=&m \bar{\sigma}^{a} \nabla_{X_{a}} \dot{\Phi}\\
&=&m^2 \Phi\\
&=&\Delta \Phi.
\ea
Here $m$ is the Majorana mass.

\smallskip
From this point on, we will consider a massive test Majorana neutrino field $\Phi$ given by the 2-component spinor
\ba
\Phi = \left ( \begin{array}{c} \varphi_{1} \\ \varphi_{2} \end{array}  \right )
\ea
satisfying the Weyl equation
\ba
\sigma^{a} \nabla_{X_{a}} \Phi =m \dot{\Phi}.
\ea

\noindent
The (odd-Grassmann) components $\varphi_{1}$ and $\varphi_{2}$
are  taken as complex valued functions of all coordinates $\{u,v,x,y\}$.
The symmetrized energy-momentum tensor can be calculated from
\ba \label{stress-energy}
T_{ab}[\Phi]  &=& \frac{i}{4} \left ( \Phi^{\dagger} \sigma_a \nabla_{X_{b}} \Phi + \Phi^{\dagger} \sigma_b \nabla_{X_{a}} \Phi
-  \nabla_{X_{a}}\Phi^{\dagger} \sigma_b \Phi - \nabla_{X_{b}}\Phi^{\dagger}  \sigma_a \Phi  \right ) \nonumber \\ 
&+&\frac{im}{2} \left ( \Psi^{\dagger}\sigma_2\Psi^* - \Psi^{T} \sigma_2 \Psi \right).
\ea

\noindent Now we examine the gravitational plane wave metric given by
\ba
g = 2du dv + F(u)^{2} dx^2 + G(u)^{2} dy^2
\ea
and derive the Weyl equation in this background spacetime. It simplifies to the following system of first-order differential equations:
\begin{eqnarray}
\left ( \frac{\partial}{\partial v} \right ) \varphi_1 +  \frac{1}{\sqrt{2}} \left ( \frac{1}{F} \frac{\partial}{\partial x}  - \frac{i}{G} \frac{\partial}{\partial y}  \right ) \varphi_2 &=& \frac{m}{\sqrt{2}}\varphi_2^*, \\
\left (-\frac{\partial}{\partial u}  -\frac{1}{2} ( \frac{F_u}{F} +  \frac{G_u}{G}  ) \right ) \varphi_2+  \frac{1}{\sqrt{2}} \left ( \frac{1}{F} \frac{\partial}{\partial x} + \frac{i}{G} \frac{\partial}{\partial y}  \right ) \varphi_1 &=& -\frac{m}{\sqrt{2}}\varphi_1^* .
\end{eqnarray}
A class of exact solutions is expressed as
\begin{eqnarray}
\varphi_1 &=& \frac{i}{p_v\sqrt{2}\sqrt{FG}}\Bigg(  \eta_1\Big(\frac{ip_1}{F}+\frac{p_2}{G}\Big)-\eta_2^*m \Bigg)e^{i(p_v + p_1 x + p_2 y)}e^{iK(u)} \nonumber \\
&+& \frac{i}{p_v\sqrt{2}\sqrt{FG}}\Bigg(  \eta_2\Big(\frac{ip_1}{F}+\frac{p_2}{G}\Big)+\eta_1^* m \Bigg)e^{-i(p_ v + p_1 x + p_2 y)}e^{-iK(u)} \nonumber\\
\varphi_2 &=& \frac{\eta_1}{\sqrt{FG}} e^{i(p_v + p_1 x + p_2 y)}e^{iK(u)}+\frac{\eta_2}{\sqrt{FG}} e^{-i(p_v + p_1 x + p_2 y)}e^{-iK(u)}.
\end{eqnarray}
\noindent
where $p_v,p_1, p_2$ represent constant momentum components, and $\eta_1,\eta_2$ are complex-valued odd-Grassmann parameters. The phase function $K(u)$ is determined separately in each region by computing the relevant integral
\ba
K(u) = -\frac{1}{2 p_v} \int_{0}^{u} \left ( \frac{p_1^2}{F(u^{\prime})^{2}} + \frac{p_2^2}{G(u^{\prime})^{2}}+m^2  \right ) du^{\prime}.
\ea
\subsection{Pure Gravitational Sandwich Wave Background}
The metric functions characterizing a purely gravitational sandwich wave are derived by setting $a=0$ in (\ref{eq9}), resulting in:
\ba
F(u)&=&\left\{
    \begin{array}{ll}
    1 , \quad u<0, & Region \hspace{2mm}I \\
    cosh(bu), \quad 0<u<u_0, & Region \hspace{2mm}I \\
    \alpha_0+\beta_0u, \quad u_0<u, & Region \hspace{2mm}III \\
\end{array}\right. \\
G(u)&=&\left\{
    \begin{array}{ll}
    1 , \quad u<0, & Region \hspace{2mm}I \\
    cos(bu), \quad  0<u<u_0, & Region \hspace{2mm}II \\
    \gamma_0-\tau_0u, \quad u_0<u, & Region \hspace{2mm}III \\
    \end{array}\right.
\ea
where
\ba
\alpha_{0}&=&cosh(bu_{0})-bu_{0}sinh(bu_{0}), \nonumber \\
\beta_{0}&=&bsinh(bu_{0}), \nonumber \\
\gamma_{0}&=&cos(bu_{0})+bu_{0}sin(bu_{0}), \nonumber \\
\tau_{0}&=&bsin(bu_{0}). \nonumber
\ea


In the solutions presented above, the phase function for each region is determined as follows
\ba
-2p_v K(u)= \left\{
    \begin{array}{ll}
        (p_1^2+p_2^2+m^2)u+c_1 & Region \hspace{2mm}I \\
        \frac{p_1^2}{b}tanh(bu)+\frac{p_2^2}{b}tan(bu)+m^2u + c_2 & Region \hspace{2mm}II \\
        \frac{p_2^2}{\tau_0(\gamma_0-\tau_0u)}-\frac{p_1^2}{\beta_0(\alpha_0+\beta_0u)} + m^2u+c_3 & Region \hspace{2mm}III, \\
    \end{array}\right.
\ea
where $c_1$, $c_2$ and $c_3$ are integration constants, which are determined by enforcing the continuity of the neutrino fields across each boundary. The computation of these constants results in
\ba
c_1=c_2=0 \quad and \quad c_3=\frac{p_1^2}{b}coth(bu_0)-\frac{p_2^2}{b}cot(bu_0).
\ea

\subsection{Pure Electromagnetic Sandwich Wave Background}
 The metric functions corresponding to a pure electromagnetic sandwich wave is obtained if we set $b=0$ in (\ref{eq9}); in this particular limit, the metric functions are
 \ba
 F(u)=G(u)=\left\{
    \begin{array}{ll}
    1 , \quad u<0, & Region \hspace{2mm}I \\
    cos(au) , \quad 0<u<u_0, & Region \hspace{2mm}II \\
    \bar{\alpha}_0-\bar{\beta}_0u , \quad u_0<u, & Region \hspace{2mm}III \\
    \end{array}\right.
 \ea
where
\ba
\bar{\alpha}_{0}&=&cos(au_{0})+au_{0}sin(au_{0}) \nonumber \\
\bar{\beta}_{0}&=&asin(au_{0}).
\ea


\noindent In the solutions above, the phase function for each region is calculated as
\ba
-2p_v K(u)= \left\{
    \begin{array}{ll}
        (p_1^2+p_2^2+m^2)u+\bar{c}_1 & Region \hspace{2mm}I \\
        \frac{p_1^2+p_2^2}{a}tan(au)+m^2u+\bar{c}_2 & Region \hspace{2mm}II \\
        \frac{p_1^2+p_2^2}{\bar{\beta}_0(\bar{\alpha}_0-\bar{\beta}_0u)}+m^2u+\bar{c}_3 & Region \hspace{2mm}III, \\
    \end{array}\right.
\ea
where $\bar{c}_1$, $\bar{c}_2$ and $\bar{c}_3$ are the integration constants to be fixed by considering the continuity of the neutrino fields across each boundary.
The calculation for the constants yields
\ba
\bar{c}_1=\bar{c}_2=0 \quad and \quad \bar{c}_3=-\frac{p_1^2+p_2^2}{a}cot(au_0).
\ea

\subsection{Mixture of gravitational and electromagnetic sandwich waves background}
The specific case of a mixture of gravitational and electromagnetic sandwich wave is obtained when $a=b$ in (\ref{eq9}), which yields
\ba
F(u)&=&1, \quad everywhere \nonumber \\
G(u)&=&\left\{
    \begin{array}{ll}
    1 , \quad u<0, & Region \hspace{2mm}I \\
    cos(\sqrt{2}au) , \quad 0<u<u_0, & Region \hspace{2mm}II \\
    \tilde{\alpha}_0-\tilde{\beta}_0u , \quad u_0<u, & Region \hspace{2mm}III \\
    \end{array}\right.
\ea
where
\ba
\tilde{\alpha}_{0}&=&cos(\sqrt{2}au_{0})+\sqrt{2}au_{0}sin(\sqrt{2}au_{0}), \nonumber \\
\tilde{\beta}_{0}&=&\sqrt{2}asin(\sqrt{2}au_{0}).
\ea


\noindent In the solutions above, the phase function for each region is calculated as
\ba
-2p_v K(u)= \left\{
    \begin{array}{ll}
        (p_1^2+p_2^2+m^2)u+\tilde{c}_1 & Region \hspace{2mm}I \\
        p_1^2u+\frac{p_2^2}{\sqrt{2}a}tan(au)+m^2u+\tilde{c}_2 & Region \hspace{2mm}II \\
        p_1^2u+\frac{p_2^2}{\tilde{\beta}_0(\tilde{\alpha}_0-\tilde{\beta}_0u)}+m^2u+\tilde{c}_3 & Region \hspace{2mm}III, \\
    \end{array}\right.
\ea
where $\tilde{c}_1$, $\tilde{c}_2$ and $\tilde{c}_3$ are the integration constants  fixed by considering the continuity of the neutrino fields across each boundary.
The calculation yields
\ba
\tilde{c}_1=\tilde{c}_2=0 \quad and \quad \tilde{c}_3=-\frac{p_2^2}{\sqrt{2}a}cot(\sqrt{2}au_0).
\ea

\section{Energy Calculations}

\noindent
To understand how a sandwich gravitational wave spacetime affects a test neutrino field, one must analyze the plane wave solutions derived earlier. A crucial step in this process is computing the components of the neutrino stress-energy-momentum tensor. The orthonormal components of $T_{ab}$ allow us to define the neutrino stress-energy 3-forms, $\tau_{a} = T_{ab} *e^b$, which can then be decomposed into a (3+1) structure for physical interpretation relative to time-like inertial observers. 
For our massive neutrino solution above, we have calculated the energy density 3-form
\ba
\rho&=&\frac{1}{\sqrt{2}p_v^2}\Bigg[(-|\eta_1|^2+|\eta_2|^2) \nonumber \\
&\times&\Big(-\big(\frac{K_u+p_v}{2}\big)\big(\frac{p_1^2}{F^2}+\frac{p_2^2}{G^2}-m^2\big)+2p_v\big(\frac{p_1^2}{F^2}+\frac{p_2^2}{G^2}\big)+K_u p_v^2+ p_v^3\Big) \nonumber \\
&-&\frac{p_1p_2}{2FG}(\frac{F_u}{F}-\frac{G_u}{G})(|\eta_1|^2+|\eta_2|^2)\Bigg]dx \w dy \w dz.
\ea
Here, we present the variation in the neutrino energy density as the incoming test neutrino transitions from Region I through Region II and into Region III.

We have in a pure gravitational sandwich wave background
\ba
\Delta \rho &=& \rho_{out} - \rho_{in} \nonumber \\
&=&\frac{\sqrt{2}}{4p_v^3}\Bigg[\left(-{| \eta_{1}|}^{2}+{| \eta_{2}|}^{2}\right) \nonumber \\
&\times& \Big(\frac{p_{1}^{4} \mathrm{sech}\! \left(b u_0 \right)^{4}}{2}+p_{1}^{2} \left(p_{2}^{2} \sec \! \left(b u_0 \right)^{2}+2 p_{v}^{2}\right) \mathrm{sech}\! \left(b u_0 \right)^{2}\nonumber \\
&+& \frac{1}{2}(p_2^2 \sec(bu_0)^2-p_1^2 -p_2^2)(p_2^2 \sec(bu_0)^2+p_1^2 +p_2^2+4p_v^2)\Big)  \nonumber \\
&-&2 \left({| \eta_{1}|}^{2}+{| \eta_{2}|}^{2}\right) b \sec \! \left(b u_0 \right) p_{1} p_{2} p_{v} \tanh \! \left(b u_0 \right) \mathrm{sech}\! \left(b u_0 \right)\Bigg],
\ea
while in a pure electromagnetic sandwich wave background,
\ba
\Delta \rho &=& \rho_{out} - \rho_{in} \nonumber \\
&=&\frac{1}{8 p_{v}^{3}}\left(-{| \eta_{1}|}^2+{| \eta_{2}|}^2\right) \nonumber\\
&\times&\left(p_{1}^{2}+p_{2}^{2}\right) \sqrt{2}\, \sec \! \left(a u_0 \right)^{2} \left(\cos \! \left(a u_0 \right)^{2} \left(p_{1}^{2}+p_{2}^{2}+4 p_{v}^{2}\right)+p_{1}^{2}+p_{2}^{2}\right) \tan \! \left(a u_0 \right)^{2}.\nonumber \\ 
\ea
On the other hand, in a mixed gravitational and electromagnetic sandwich wave background we have
\ba
\Delta \rho &=& \rho_{out} - \rho_{in} \nonumber \\
&=&\frac{\sqrt{2}}{4p_v^3}\Bigg[\left(-{| \eta_{1}|}^{2}+{| \eta_{2}|}^{2}\right) \nonumber \\ 
&\times&\left(\frac{p_{2}^{4} \sec \! \left(\sqrt{2}\, a u_0 \right)^{4}}{2}+p_{2}^{2} \left(p_{1}^{2}+2 p_{v}^{2}\right) \sec \! \left(\sqrt{2}\, a u_0 \right)^{2}-\left(p_{1}^{2}+\frac{p_{2}^{2}}{2}+2 p_{v}^{2}\right) p_{2}^{2}\right) \nonumber \\
&-&\left({| \eta_{1}|}^{2}+{| \eta_{2}|}^{2}\right) \sqrt{2}a \tan \! \left(\sqrt{2}\, a u_0 \right) p_{1} p_{v} p_{2} \Bigg].
\ea

\bigskip

\section{Conclusion}
In this work, we analyzed the behavior of a test neutrino field in the background of a gravitational sandwich wave spacetime. The metric is expressed in Rosen coordinates $(u,v,x,y)$.
A sandwich wave is a particular type of gravitational wave in which the curvature is non-vanishing only within a finite interval $0 \leq u \leq u_0$.
We considered three separate cases: a purely gravitational sandwich wave, a purely electromagnetic sandwich wave, and a mixed gravitational-electromagnetic sandwich wave. In each scenario, the spacetime reduces to the flat Minkowski metric before and after the wave zone.
We obtained exact background solutions of the Weyl equation in all three regions and matched them across the wave boundaries using the O'Brien-Synge junction conditions. Our results demonstrate that the passage of a neutrino through a sandwich gravitational wave leads to variations in its energy density. Depending on the parameters, the test neutrino may experience oscillations in its energy as well as a phase shift in its wave function. Both effects are noteworthy in their own right and merit further investigation, particularly with regard to their potential observability today.

\noindent Although the standard electroweak theory originally described three neutrino species as massless and left-chiral, it is now well established that neutrinos can transform from one flavor into another during propagation. These flavor oscillations require nonzero neutrino masses with finite mass differences between species. To fully capture such effects, one must extend the analysis to include at least two neutrino species, an avenue we leave for future work. In our study, we have shown that test neutrinos can exchange energy with sandwich gravitational waves. This suggests that, in a more complete framework involving multiple neutrino species, gravitational interactions may play a role in shaping neutrino mass differences, thereby influencing or even inducing flavor oscillations.

\bigskip

\section{Acknowledgement}
One of us (T.D.) thanks the Turkish Academy of Sciences (TUBA) for partial support.

\end{document}